\documentclass[a4paper,10pt,twoside]{cpc-hepnp}
\usepackage{CJK,upgreek,fancyhdr}
\usepackage{multicol}
\usepackage{graphicx}
\usepackage{booktabs}
\usepackage{amssymb,bm,mathrsfs,bbm,amscd}
\usepackage[tbtags]{amsmath}
\usepackage{lastpage}
\usepackage{multirow}
\usepackage{caption}
\usepackage{float}

\begin{document}
\begin{CJK*}{GB}{gbsn}

\fancyhead[c]{\small Chinese Physics C~~~Vol. xx, No. x (202x) xxxxxx}
\fancyfoot[C]{\small 010201-\thepage}

\footnotetext[0]{Received 5 March 2026}

\title{Study of radiative proton capture by the $^{7}$Be nucleus with the use of ab initio approaches}

\author{%
      D. M. Rodkin $^{1,2,3;1)}$\email{rodkindm92@gmail.com}%
\quad Yu. M. Tchuvil'sky $^{1,2;2)}$\email{tchuvlyuri@gmail.com}%
}
\maketitle

\address{%
$^1$ Dukhov Research Institute for Automatics, 127055, Moscow, Russia; \\
$^2$ Skobeltsyn Institute of Nuclear Physics, Lomonosov Moscow State University, 119991 Moscow, Russia;\\
$^3$ RUDN University, 6 Miklukho-Maklaya Street,117198 Moscow, Russia.\\
}

\begin{abstract}
A theoretical study of the $^7$Be(p,$\gamma$)$^8$B reaction in the "astrophysical" energy range with the use of ab initio methods is presented. The used approaches are No-Core Shell Model and Cluster Channels Orthogonal Functions Method. The scheme also contains elements of R-matrix theory and procedures for extrapolating various data obtained in ab initio computations. The developed approach as a whole allows one not only to calculate the astrophysical S-factor and all nuclear characteristics that determine its value, but also to evaluate the reliability of the obtained results and to identify the dominant reaction mechanisms against a background of insignificant ones.The high accuracy of the obtained results and has been demonstrated.
\end{abstract}

\begin{keyword}
no-core shell model, realistic NN-potentials, asymptotic properties of bound and resonance states, radiative capture reactions.
\end{keyword}
\begin{pacs}
03.65.-w, 03.65.Nk, 03.54.Fd
\end{pacs}

\footnotetext[0]{\hspace*{-3mm}\raisebox{0.3ex}{$\scriptstyle\copyright$}2013
Chinese Physical Society and the Institute of High Energy Physics
of the Chinese Academy of Sciences and the Institute
of Modern Physics of the Chinese Academy of Sciences and IOP Publishing Ltd}

\begin{multicols}{2}

\section{Introduction}

The reaction of proton radiative capture by $^7$Be is the essential part of the pp-chain or more precisely -- pp-III chain \cite{pgamma}. The radiative capture reaction $^7$Be(p, $\gamma$)$^8$B played an important role in uncovering the apparent loss of solar neutrino flux. The flux of $^8$B neutrinos, detected by the Super-Kamiokande \cite{sk} and Sudbury Neutrino Observatory \cite{sno}, contrasted with predictions from standard solar model(SSM), led to the discovery of the neutrino oscillations. For the current tasks of neutrino physics cross-section of $^7$Be(p, $\gamma$)$^8$B reaction should be well studied around $\approx$ 20 keV incident proton energy, but experiments in the low-energy region are difficult and currently limited to energies above 100 keV. Thus, the required cross-section can come from low-energy extrapolations of data or from direct computations.

For available energies this reaction is rather well studied -- \cite{exp1, exp2, exp3, exp4, exp5,exp6,exp7}. Data from these experiments agree well, so it looks optimal to compare the results of computations presented bellow with the latest known experimental data \cite{exp6}, as one of the most complete. 

On the other hand, the results of the extrapolation procedures for the low energy region differ quite significantly \cite{exp7, extrap} - the S$_{17}$(0) varies from 17.1 $\mp$ 0.5 to 22.6 $\mp$ 0.2 and the most accurate value is assumed to be 20.9 $\mp$ 0.6 (exp) $\mp$ 0.7(theor.) eV  $\cdot$ Bn. The alternative way of obtaining S$_{17}$(0) is by extracting asymptotic normalization coefficients (ANCs) from breakup reactions $^{10}$B($^7$Be, $^8$B)$^9$Be, $^{14}$N($^7$Be, $^8$B)$^{13}$C and $^{13}$C($^7$Li, $^8$Li)$^{12}$C \cite{breakup}. The extracted squared ANC is $C^{2}_{tot} = 0.450 \mp 0.039$ fm$^{-1}$, that leads to the astrophysical factor $S_{17} (0) = 17.6 \mp 1.7$ eV $\cdot$ Bn. 

So, the discrepancy in the results of analysis of experimental data leads to necessity of the use of various theoretical approaches. 

Modern high-precision methods for describing both light nuclei properties and characteristics of reactions induced by light nuclei collisions are advancing nowadays. An important role among the methods describing light nuclei structure belongs to various ab initio methods. These approaches are based on new possibilities provided by modern high-performance supercomputers and on the use of realistic nucleon-nucleon potentials. These potentials could be derived from Chiral Effective Field Theory \cite{cheft1, cheft2, cheft3} or from nucleon scattering data by the use of J-matrix inverse scattering method \cite{jisp1}. In the current work the Daejeon16 NN-potential \cite{cheft1} which is built using the N$^3$LO limitation of Chiral Effective Field Theory is exploited.

The most popular among ab initio methods describing nuclear structure are various versions of No-Core Shell Model (NCSM) \cite{ncsm1}, Gamov Shell Model (GSM) \cite{gsm1}, Green functions Monte Carlo method \cite{gfmc1} and the Coupled Cluster Method \cite{ccm1}.

NCSM model and methods similar to it are, however, not adapted to describe clustering effects and, consequently, wave functions of channels of nuclear reactions, as well as decay properties of resonances directly. For this purposes, different methods were developed. Among them there are methods which combine NCSM and RGM namely No-Core Shell Model / Resonating Group Model (NCSM/RGM) \cite{ncsmrgm1} and No-Core Shell Model with Continuum (NCSMC) \cite{ncsmc1} are considered the most developed. As the NCSMC the Fermionic Molecular Dynamics (FMD) \cite{fmd1} offers in fact an ab initio approach focused on the unified description of both bound states and continuum ones.

For theoretical studies of nuclear reactions on light nuclei many approaches which are not ab initio ones are used. The use of phenomenological potentials together with truncated bases leads to simplifying calculations and, in some cases, better description of long-range asymptotic. Among these approaches one can mention Generator Coordinate Method (GCM) \cite{gcm1}, Microscopic Cluster Model \cite{mcm1}, Antisymmetrized Molecular Dynamics (AMD) \cite{amd1}, Algebraic Version of the RGM \cite{avrgm1}, Coupled Cluster Method \cite{ccm1} and Coupled Cluster Gamov Shell Model - (GSM-CC) \cite{cc-gsm1}. Hartree-Fock approach with the Skyrme forces is also used for studying the discussed reactions \cite{hfb1}.

Both ab initio methods and methods using effective nucleon-nucleon potentials were used for studying $^7$Be(p,$\gamma$)$^8$B reaction.  Most of these calculations turn out to be in good agreement with the pattern of experimental values of radiative proton capture cross-sections in the energy area, available for the measurements. However, they show rather different values of astrophysical S-factor at zero energy $S_{17} (0)$. Indeed, for Hartree-Fock approach with the Skyrme forces \cite{hfb1} this value is $S_{17} (0) = 22.3$ eV $\cdot$ Bn. In case of the use of coupled-channel formalism \cite{ccm2, ccm3} $S_{17}(0)$ are 20.9 and 21.0, respectively. The GSM-CC model was also used for $^7$Be(p,$\gamma$)$^8$B process calculation with the use of an effective finite-range two-body interaction and inert $^4$He core -- $S_{17}(0) = 23.2143??$ eV $\cdot$ Bn \cite{cc-gsm1}. The most theoretically justified results were obtained by the use of NCSMC method \cite{ncsmc2}. These calculations suggest the value for the $^7$Be(p,$\gamma$)$^8$B S-factor at zero energy of 19.8 $\pm$ 0.3 eV $\cdot$ Bn for N$^4$LO interaction and 21.0 eV $\cdot$ Bn for N$^3$LO interaction.

Thus, the above presented results of calculations demonstrate an essential scatter at low energies. Such a situation confirms the importance of further theoretical investigations, especially using well-founded ab initio methods. In the current work the results which have been obtained in the framework of approach of such a type, namely Cluster Channel Orthogonal Functions Method (CCOFM) are presented. We have previously demonstrated the high quality of the description by this method of the decay widths and ANCs of various nucleon and cluster channels \cite{our0}, which gives hope for its successful application in both the discussed and many other problems of nuclear astrophysics. One of the basic advantages of this scheme is that, in contrast to the previously developed ones, it allows one to calculate not only total decay widths but the reduced partial width amplitudes (RPWAs) of resonance states into various cluster channels simultaneously. The method is based on the employment of NCSM computations. The possibility to calculate ANCs and RPWAs of decay channels within CCOFM makes it possible to develop the method of using ab initio calculated quantities in the theoretical studies of resonance nuclear reactions because the cross-sections of these reactions in the R-matrix theory are expressed in terms of ANCs and RPWAs.

Earlier this method was successfully used for studying elastic scattering of neutrons on $^9$Be \cite{our2} and for obtaining the cross-sections of resonance nuclear reactions going throw $^8$Be compound nucleus -- \cite{our3, our4}.

To study the proton radiative capture process, the CCOFM approach must be supplemented by NCSM calculations of the electromagnetic transition amplitudes. Methods allow one to increase the quality of calculations of such values are present in the present paper.  

\section{Formalism of calculating spectroscopic factors, cluster form factors, asymptotic properties and widths of electromagnetic transitions}

Let us demonstrate how translationally-invariant A-nucleon wave functions (WFs) of arbitrary two-fragment decay channel with separation A = A$_1$ + A$_2$ are built in CCOFM. The useful feature of this procedure is that each function of this basis can be represented as a superposition of Slater determinants (SDs). To do that the technique of so-called cluster coefficients (CCs) is exploited. 

The oscillator-basis terms of the cluster channel $c_\kappa$ are expressed in the following form:
\begin{equation}
\Psi^{c_\kappa} _{{A\,},nlm}  = \hat A\{\Psi^{\{k_1\}}_{A\,_1 } \Psi^{\{k_2\}}_{A\,_2 }
\varphi _{nlm} (\vec \rho )\}_{J_cJM_JT} , 
\label{eq1}
\end{equation}
where  $\hat A$ is the antisymmetrizer, $\Psi^{\{k_i\}}_{A\,_i}$ is a translationally-invariant internal WF of the fragment labelled by a set of quantum numbers $\{k_i\}$; $\varphi _{nlm} (\vec \rho )$ is the WF of the relative motion. The channel WF is labelled by the set of quantum numbers $c_\kappa$ which includes $\{k_1\},\{k_2\},n,l,J_c,J,M_J,T$, where $J$ is the total momentum and $J_c$ is the channel spin.

The basic idea of the method is to represent each function of the cluster basis
as a linear combination of the functions of the M-scheme.  To do that function (\ref{eq1})
is multiplied by the function of the center of mass (CM) zero vibrations $\Phi _{000}
(\vec R)$. Then the transformation of WFs caused by changing from $\vec R,\vec \rho$ to
$\vec R_1 ,\vec R_2$ coordinates -- different-mass Talmi-Moshinsky-Smirnov transformation \cite{talmi-moshinsky} -- is performed and WF (\ref{eq1}) takes the form

\begin{equation}
\begin{array}{rcl}
\Phi _{000} (\vec R)\Psi^{c_\kappa} _{{A\,},nlm} = \sum\limits_{N_i ,L_i ,M_i}
{\left\langle {{\begin{array}{*{20}c}
   {000}  \\
   {nlm}  \\
\end{array}  }}
 \mathrel{\left | {\vphantom {{\begin{array}{*{20}c}
   {000}  \\
   {nlm}  \\
\end{array}  } {\begin{array}{*{20}c}
   {N_1 ,L_1 ,M_1 }  \\
   {N_2 ,L_2 ,M_2 }  \\
\end{array}}}}
 \right. \kern-\nulldelimiterspace}
 {{\begin{array}{*{20}c}
   {N_1 ,L_1 ,M_1 }  \\
   {N_2 ,L_2 ,M_2 }  \\
\end{array}}} \right\rangle } \\[\bigskipamount]
\hat A\{ \Phi _{N_1 ,L_1 ,M_1 }^{A_1 } (\vec R_1 ) \Psi^{\{k_1\}}_{A\,_1 } \Phi _{N_2
,L_2 ,M_2 }^{A_2 } (\vec R_2 )\Psi^{\{k_2\}} _{A\,_2 } \} _{J_c,M_{J_c},M_JT} . \label{eq2}
\end{array}
\end{equation}

The key technical procedure of the method is to transform each of the two products of the internal WFs of the fragments with their functions of non-zero center-of-mass (CM) oscillations into a superposition of Slater determinants (SDs)

\begin{equation}
\Phi _{N_i ,L_i ,M_i }^{A_i } (\vec R_i )\Psi^{\{k_i\}}_{A\,_i }  = \sum\limits_k
{X_{N_i ,L_i ,M_i }^{A_i (k)} \Psi _{A\,_i (k)}^{SD} }. \label{eq3}
\end{equation}
Quantity $X_{N_i ,L_i ,M_i }^{A_i (k)}$ is called a cluster coefficient (CC). Technique of
these objects are presented in detail in \cite{nem}. In this work the formalism based on the method of the second quantization of the oscillator quanta is used. It was described in details in \cite{our5}. As a result of these transformations, the antisymmetrized product of functions in the right-hand side of expression (\ref{eq2}) also turns out to be a superposition of SDs.

It should be noted that WFs of cluster-channel basis terms (\ref{eq1}) of one and the same
channel $c_\kappa$ characterized by the pair of internal functions
$\Psi^{\{k_1\}}_{A_1}$, $\Psi^{\{k_2\}}_{A_2}$  and extra quantum numbers
$l,J_c,J,M_J,T$ briefly denoted as $\Psi^{c_\kappa} _{{A\,},n}$ are non-normalized due to the properties of the antisymmetrization operator and, with rare exceptions, non-orthogonal. Creation of orthonormalized basis functions of a separate channel $c_\kappa$ is performed by the diagonalization of the overlap kernel matrix

$$ ||N_{nn'} || \equiv  \langle \Psi^{c_\kappa} _{{A\,},n'}|\Psi^{c_\kappa} _{{A\,},n} \rangle = $$
\begin{equation}
\langle \Psi^{\{k_1\}} _{A_1} \,\Psi^{\{k_2\}} _{A_2} \,\varphi _{nl} (\rho )
|\hat A^2 |\Psi^{\{k_1\}} _{A_1} \,\Psi^{\{k_2\}} _{A_2} \,\varphi _{n'l} (\rho )\Phi _{00}(\vec R) \rangle . \label{eq4}
\end{equation}

The eigenvalues and eigenvectors of this overlap kernel are the same in the shell-model and translationally-invariant representations and can be written as:

\begin{equation}
\varepsilon _{\kappa,k}  =  \langle \hat A\{ \Psi^{\{k_1\}} _{A_1} \,\Psi^{\{k_2\}} _{A_2} \,f_l^k (\rho
) \} |\hat 1|\hat A\{ \Psi^{\{k_1\}} _{A_1} \,\Psi^{\{k_2\}}_{A_2} \,f_l^k (\rho
) \}  \rangle ; \label{eq5}
\end{equation}

\begin{equation}
f_l^k (\rho ) = \sum\limits_n {B_{nl}^k \varphi _{nl} (\rho )}. \label{eq6}
\end{equation}

On the other hand, the WFs of the orthonormalized channel basis $c_\kappa$ 

\begin{equation}
\Psi^{SD,c_\kappa} _{{A\,},kl}=\varepsilon^{-1/2} _{\kappa,k}|\Phi _{00}(\vec R)\hat A\{ \Psi^{\{k_1\}}
_{A_1} \,\Psi^{\{k_2\}}_{A_2} \,f_l^k (\rho ) \}  \rangle. \label{eq7}
\end{equation}
turn out to be represented in the form of the superposition of the SDs. The basis of such functions is complete in the sense that a function of this channel

\begin{equation}
\Psi^{SD,c_\kappa} = \Phi _{00}(\vec R)\hat A\{\Psi^{\{k_1\}} _{A\,_1} \Psi^{\{k_2\}}
_{A\,_2}\Phi(\rho) Y_{l} (\Omega)\} _{J_c,M_JT}  \label{eq8}
\end{equation}
including arbitrary WF $\Phi(\rho)$ can be represented as a superposition of such WFs.

The cluster form factor (CFF) $\Phi^{c_\kappa}_A(\rho)$  describes the relative motion of subsystems in A-nucleon configuration space and is defined by the following overlap
$$\Phi^{c_\kappa}_A(\rho)=$$
\begin{equation}
\langle \Psi _{A}|\hat N^{-1/2}\hat A\{\Psi^{\{k_1\}} _{A\,_1} \Psi^{\{k_2\}}
_{A\,_2}\frac{\delta(\rho-\rho')}{\rho'^2}Y_{l} (\Omega)\} _{J_c,M_JT}\rangle ,\label{eq9}
\end{equation} 
where $\Psi _{A}$  is the WF of the initial nucleus and $\hat N$ is the exchange kernel operator  which takes the form:

$$\hat N(\rho', \rho'')=\langle \hat A\{\Psi^{\{k_1\}} _{A\,_1} \Psi^{\{k_2\}}
_{A\,_2}\frac{\delta(\rho-\rho')}{\rho'^2}Y_{l} (\Omega)\} _{J_c,J,M_J,T}|\times$$
\begin{equation}
|\hat A\{\Psi^{\{k_1\}} _{A\,_1} \Psi^{\{k_2\}}
_{A\,_2}\frac{\delta(\rho-\rho'')}{\rho''^2}Y_{l} (\Omega)\} _{J_c,J,M_J,T}\rangle.\label{eq10}
\end{equation} 
Representation of the generalized function of the relative motion in the form of an expansion in terms of oscillator functions

\begin{equation}
[\delta(\rho-\rho')/\rho'^2]Y_{lm} (\Omega)=\sum\limits_n\varphi _{nlm} (\vec \rho )\varphi _{nlm} (\vec \rho^{'} )\label{eq11}
\end{equation}
first, reduces the exchange kernel operator to the overlap kernel matrix (\ref{eq4}) and, second, makes it possible to write the CFF in the form 
 
\begin{equation}
\Phi^{c_\kappa}_A(\rho)=\sum\limits_k \varepsilon^{-1/2} _{\kappa,k} \langle \Psi _{A}|\hat A\{ \Psi^{\{k_1\}} _{A_1}
\,\Psi^{\{k_2\}}_{A_2} \,f_l^k (\rho') \} \rangle f_l^k (\rho ).\label{eq12}
\end{equation}
 
 After that the CFF can be expressed in the form of an expansion in the oscillator basis using above presented techniques:
\begin{equation}
\Phi^{c_\kappa}_A(\rho)=\sum\limits_k\varepsilon^{-1/2} _{\kappa,k}\sum\limits_{n, n'}B_{nl}^k B_{n'l}^{k}
C_{AA_1A_2}^{n'l}\varphi _{nl} (\rho ) \label{eq13}
\end{equation}
 The coefficients $C_{AA_1A_2}^{nl}$ contained in this expression have the form
$$C_{AA_1A_2}^{nl} = \langle \hat A\{\Psi^{\{k_1\}} _{A_1} \Psi^{\{k_2\}} _{A_2} \varphi _{nl}
(\rho ) \} | \Psi _{A} \rangle =$$
\begin{equation}
\langle \Psi^{SD,c_\kappa} _{A, nl}|\Phi _{000} (R)| \Psi _{A} \rangle = \langle \Psi^{SD,c_\kappa} _{A,
nl}|\Psi^{SM} _{A} \rangle. \label{eq14}
\end{equation}
These coefficients are traditionally called the spectroscopic amplitudes (SAs).

The spectroscopic factor (SF) is defined as the norm of CFF, for the discussed channel $c_\kappa$it can be written as
$$S_l^{c_\kappa}  = \int {|\Phi^{c_\kappa}_A(\rho)|^2 } \rho ^2 d\rho =$$
\begin{equation}
\sum\limits_k {\varepsilon _k^{ - 1} \sum\limits_{nn'} {C_{AA_1A_2}^{nl} }
C_{AA_1A_2}^{n'l} B_{nl}^k } \;B_{n'l}^k . \label{eq15}
\end{equation}

Both SFs and CFFs are the objects used in theoretical studies on nuclear decays and reactions. Evidently CFF is more informative characteristic of a cluster channel. In CCOFM the obtained CFFs are exploited for computing the widths of resonances. The norms of these values -- SFs -- are used to distinguish the main channels against the background of a multitude of other ones, existence of which practically do not affect the results of experiments. 

We use the procedure of logarithmic matching of the CFF with the asymptotic WF of the corresponding open decay channel:  

\begin{equation}
\Xi_l ( \rho)  = \frac{(F^2_l (\eta, k \rho) + G^2_l (\eta, k \rho))^{1/2}}{k \rho}. \label{eq16}
\end{equation}

 The partial decay channel width is obtained within the expression relative to one used in traditional R-matrix theory:
\begin{equation}
\Gamma  = \frac{\hbar^2}{\mu k_0} \Xi_l ( \rho_{m})^{-2}(\Phi _A^{c_\kappa} (\rho_{m} ))^2, \label{eq17}
\end{equation}
where $\rho_m$ -- the matching point of logarithmic derivatives.
The amplitude of reduced partial decay width is expressed throw the value of CFF in the matching point as
\begin{equation}
\gamma_{c_{\kappa}} = (\frac{\hbar^2 \rho_m}{2 \mu_c})^{1/2} \Phi_A^{c_{\kappa}}(\rho_m).\label{eq18}
\end{equation}	

In R-matrix calculations the sign of partial decay width amplitude is needed, which is determined by sign of CFF in the matching point $\Phi_A^{c_{\kappa}}(\rho_m)$.

The asymptotic normalization coefficient $A^{c_\kappa}$ is determined by the same procedure by comparison of CFF with the Whittaker function $W_{-\eta,l+1/2}(2k\rho)$ in the same range of distances

\begin{equation}
A^{c_\kappa}= \rho_{m} \Phi _A^{c_\kappa} (\rho_{m} )/W_{-\eta,l+1/2}(2k \rho_{m}). \label{eq19}
\end{equation}
 
The real possibility to calculate ANCs and reduced partial width amplitudes (RPWAs) of decay channels within the CCOFM makes it possible to develop the method of using ab initio calculated quantities in the theory of nuclear reactions because the amplitudes and cross-sections of resonant reactions in the R-matrix theory are also expressed in terms of ANCs and RPWAs.

There are many computer implementations of R-matrix theory. The formulated task is most appropriately completed using the AZURE2 R-matrix code \cite{azure}. 

In the current work, the WFs of $^7$Be and $^8$B nuclei are calculated within the standard M-scheme of the NCSM on the complete basis of Slater determinants with the cut-off parameter N$^*_{max}$ in the space of total number of excitation quanta. NCSM calculations are carried out with the use of well-known shell-model code Bigstick \cite{bigstick}.

But for nuclei around A $\approx$ 8 NCSM calculations are not fully converged even for maximum allowed cut-off parameter. The only way out of this situation is the use of one of the extrapolation procedures. We use the five-parameter "Extrapolation A5" method \cite{A5} when needed.The extrapolation function depends on the five free parameters: $E_{\infty}, a, c, d, k_{\infty}$ and also on parameters $b = \sqrt{{\hbar/}{m \omega}}$, $\Lambda_{i} = b^{-1}\sqrt{2(N^{*max}_{tot,i} + 3/2)}$, $L_{i} = b\sqrt{2 (N^{*max}_{tot,i} + 3/2)}$, $L_t = L_i + 0.54437b(L_{i=0}/ b)^{-1/3}$. This function is written in the following form:
   
\begin{equation}
E_{state}(N^{*max}_{tot,i}, \hbar \omega) = E_{\infty} + a \cdot exp(-c \Lambda_{i}^2) + d \cdot exp(-2 k_{\infty} L_t) \label{eq20}
\end{equation}

The values of free parameters are determined for each level independently by fitting this function to theoretically calculated total binding energies (TBE).

In our previous paper \cite{our2} it was shown that for deeply sub-barrier resonances, even a moderate deviation of the calculation resonance energy from the experimental one often leads to a dramatic, sometimes by several orders of magnitude, changes in the calculated values of cross-sections of resonance nuclear reactions. On the other hand, the accuracy of NCSM calculations is limited even after the use of an extrapolation procedure. This reason necessitates to introduce the experimental values of resonance energies to the computation of decay widths. Note that the theoretical groups performing similar ab initio studies of the decay processes also use the same procedure in certain situations. It has become known as NCSMC-pheno in the literature.\cite{ncsmc-pheno}

The electromagnetic decay widths could be calculated with the use of NCSM model. The possibility of both EJ and MJ transitions are expressed throw reduced transition probabilities:

\begin{equation}
P(EJ, MJ) = 8 \pi \frac{e^2}{\hbar} \frac{J+1}{J[(2J+1)!!]^2} k^{2J+1} B(EJ, MJ). \label{eq21}
\end{equation}

The reduced transition probabilities B(EJ, MJ) is defined in terms of reduced matrix elements of a one-body operators by:

\begin{equation}
B(i \rightarrow f) = \frac{|\langle J_f || O(\lambda) || J_i \rangle|^2}{(2J_i + 1)}. \label{eq22}
\end{equation}

The one-body operators $O(\lambda)$ represent a sum over the operators for the individual
nucleon degrees of freedom $i$

\begin{equation}
O(\lambda) = \sum_{i} O(\lambda, i). \label{eq23}
\end{equation}

For electric transition one-body operator is defined as

\begin{equation}
O(E\lambda) = r^{\lambda} Y^{\lambda}_{\mu} (\hat r) e_q e, \label{eq24}
\end{equation} 
and for magnetic transitions operator is defined as

\begin{equation}
O(M\lambda) = \left[ \vec l \frac{2g_q^l}{\lambda + 1} + \vec s g^s_q \right] \vec \nabla \left[ r^{\lambda} Y^{\lambda}_{\mu} (\hat r) \right] \mu_{N}.  \label{eq25}
\end{equation} 

Using the Bigstick shell model code it one can calculate the reduced matrix elements as a sum over one-body transition densities times single-particle matrix elements:
\begin{equation}
\langle f || O(\lambda) || i \rangle = \sum_{k_{\alpha}, k_{\beta}} OBTD(fik_{\alpha}k_{\beta}\lambda) \langle k_{\alpha} || O(\lambda) || k_{\beta} \rangle \label{eq26}
\end{equation}
where OBTD is given by
\begin{equation}
OBTD(fik_{\alpha}k_{\beta}\lambda) = \frac{\langle f || \left[ a^{+}_{k_{\alpha}} \cdot \tilde a_{k_{\beta}} \right] ||i \rangle}{\sqrt(2\lambda + 1)}. \label{eq27}
\end{equation}

Similarly to partial decay widths for resonances located deep below the barrier electromagnetic transitions probabilities depend strongly on resonance energies. On the other hand, the accuracy of NCSM calculations is limited to $\approx$ 100 keV even for ground states. So, one should also introduce the experimental values of resonance energies (if they are known) to the computation of electromagnetic transitions.

\section{Results of calculations of the properties of $^7$Be and $^8$B states, cross-section and astrophysical S-factor of proton radiative capture by $^{7}$Be}

First of all, let us consider the properties of nuclear states that can affect the astrophysical S-factor of reaction $^7$Be(p,$\gamma$)$^8$B at low energies.

The NCSM computations of TBEs and WFs of $^{7}$Be and $^8$B nuclei were carried out in wide range of oscillator parameter $\hbar \omega$ = 10$\div$25 MeV and the basis cut-off parameter N$^{*}_{max}$ = 4$\div$12. The maximal basis for $^7$Be contain 2.52 $\cdot 10^{8}$ SDs and for $^8$B -- 9.46 $\cdot 10^{8}$ SDs. Table 1 demonstrates the pattern of convergence of NCSM calculation results for ground state of $^8$B. Extrapolation procedure A5 (\ref{eq20}) results in the TBE value -38.116 $\pm$ 0.118 i. e. the accuracy of the NCSM computations even on the great basis is about 100 keV. It should be stressed that this inaccuracy does not include possible shortcomings of the using NN-potential.

\begin{flushleft}
\tabcaption{TBEs [MeV] of ground state 2$^{+}$ of $^8$B for various values of $\hbar \omega$ and cut-off parameter $N^{*}_{max}$ and extrapolated ones.}
\begin{tabular}{l l l l l l }
\hline\hline\noalign{\smallskip}
$\hbar \omega / N^*_{max}$ & 4 & 6 & 8 & 10 & 12 \\
\hline\noalign{\smallskip} 
 10.0 & -29.02 & -33.02 & -35.43 & -36.78 & -37.50   \\
\noalign{\smallskip}
 12.5 & -32.89 & -35.70 & -37.06 & -37.67 & -37.95   \\
\noalign{\smallskip}
 15.0 & -34.44 & -36.45 & -37.34 & -37.75 & -37.95   \\
\noalign{\smallskip}
17.5 & -34.57 & -36.35 & -37.17 & -37.59 & -37.82   \\
\noalign{\smallskip}
20.0 & -33.88 & -35.81 & -36.76 & -37.29 & -37.59   \\
\hline\noalign{\smallskip}
\end{tabular}
\end{flushleft}

In total the TBEs were computed for the ground state of $^7$Be -- 3/2$^-_1$ and five lowest states of $^8$B properties of which may have a significant impact on the process under study: 2$_1^+$, 1$_1^+$, 3$_1^+$, 0$_1^+$ and 1$_2^+$. The results of calculations and the corresponding experimental data are demonstrated in Table 2. 
Computations of $^7$Be ground state and lowest states of $^8$B show overestimation in TBE in comparison with experimental data which is approximately 300 keV. The proton channel energies, being differential quantities, are reproduced with the accuracy about few tens keV. An exception is observed for 1$_1^+$ resonance. No doubt that this difference is a consequence of the imperfection of Daejeon16 potential

\begin{flushleft}
\tabcaption{Computed TBEs [MeV] of $^7$Be and $^8$B states, decay energies, and their experimental values.}
\begin{tabular*}{0.39\textwidth}{c c c c c}
\hline\noalign{\smallskip}
J$^{\pi}$ & TBE$_{th.}$ &  TBE$_{exp.}$ & E$_{p}^{th.}$ & E$_{p}^{exp.}$  \\
\hline\noalign{\smallskip} 
 \multicolumn{5}{c}{Ground state of $^7$Be.}   \\
\noalign{\smallskip}
3/2$^-_1$ & -37.965 & -37.600 & --- & ----  \\
\noalign{\smallskip}
 \multicolumn{5}{c}{Lowest states of $^8$B.}   \\
\noalign{\smallskip}
2$_1^{+}$ & -38.116 & -37.737 & -0.151 & -0.1375  \\
\noalign{\smallskip}
1$_1^+$ & -36.875 & -36.968 & 1.090 & 0.632 \\
\noalign{\smallskip}
3$_1^+$ & -35.799 & -35.417 & 2.165 & 2.182  \\
\noalign{\smallskip}
0$_1^+$ & -35.381 & --- & 2.583 & ---  \\
\noalign{\smallskip}
1$_2^+$ & -34.663 & --- & 3.301 & ---  \\
\hline\noalign{\smallskip}
\end{tabular*}
\end{flushleft}

The ANCs, widths and transition probabilities are very sensitive to the energy of a process therefore the use of well-measured level energies is preferable in the computations of the cross-sections of resonance reactions. So, we use them in $^7$Be(p,$\gamma$)$^8$B reaction cross-section calculation. 

Since we calculate the cross-section of proton radiative capture only up to the proton energy of 2.5 MeV, the impact of 1$_2^+$ state obtained in the performed studies can be neglected. Furthermore, the 0$_1^+$ resonance also practically does not contribute to this reaction cross-section, since only a low-intensity E2 transition to the 2$_1^+$ ground state is allowed for it.  

The impact of 1$_1^+$ and 3$_1^+$ resonances on the cross-section of $^7$Be(p,$\gamma$)$^8$B process is determined mainly by width of M1 transitions. It is clear a priory and has been confirmed (see bellow), that the impact of E2 transitions can be neglected. The values of reduced transition probabilities B($i \rightarrow f$) and electromagnetic widths for $3_1^{+} \rightarrow 2_1^+$ and $1_1^{+} \rightarrow 2_1^+$ M1 transitions for experimental resonance energies are shown in Table 3.

As it can be seen in Table 3 the reduced probability of $B(1_1^{+} \rightarrow 2_1^+)$ process converges well -- its computed value lies within the range $3.7675 \pm 0.0025$. Thus, the width of electromagnetic transition $1_1^{+} \rightarrow 2_1^+$ is (2.86 $\pm$ 0.006)$\cdot$ 10$^{-2}$ eV. This result is in very good agreement with experimental data from \cite{tilley-8} -- $(2.52 \pm 0.11) \cdot 10^{-2}$ eV.  

The computation of electromagnetic transition widths $3_1^{+} \rightarrow 2_1^+$ show it is $0.10 \pm 0.0036$ eV. It is also in rather good agreement with experimental value $0.1 \pm 0.05$ eV presented in \cite{tilley-8}. 

The convergence of calculations of E2 transitions is much worse. Nevertheless, there is no obstacle to obtain more or less actual estimate of the widths of $1_1^{+} \rightarrow 2_1^+$ and $3_1^{+} \rightarrow 2_1^+$ E2 transitions. For $1_1^{+} \rightarrow 2_1^+$ E2 transition the result is (3.485 $\pm$ 0.437$\mu_N^2$) $\cdot$ 10$^{-6}$ eV and for $3_1^{+} \rightarrow 2_1^+$ the width is (7.637 $\pm$ 0.257) $\cdot$ 10$^{-4}$ eV. These estimates reliably confirm the priory assumption that the contribution of E2 transitions amplitudes to the cross-section of $^7$Be(p, $\gamma$)$^8$B reaction can be neglected.

So, the convergence of the results of calculating the total widths of the electromagnetic transitions is beyond doubt.

\begin{flushleft}
\tabcaption{The reduced probabilities $B(i \rightarrow f)$ [$\mu_N^2$] and widths [eV] of $3_1^{+} \rightarrow 2_1^+$ and $1_1^{+} \rightarrow 2_1^+$ M1 transitions.}
\begin{tabular*}{0.5\textwidth}{c c c c c c}
\hline\noalign{\smallskip}
$\hbar \omega / N^*_{max}$ & 4 & 6 & 8 & 10 & 12 \\
\hline\noalign{\smallskip} 
 \multicolumn{6}{c}{Reduced probabilities of $B(1_1^{+} \rightarrow 2_1^+)$.}   \\
\noalign{\smallskip}
10.0 & 3.609 & 3.678 & 3.731 & 3.751 & 3.765  \\
\noalign{\smallskip}
15.0 & 3.902 & 3.813 & 3.791 & 3.772 & 3.770  \\
\noalign{\smallskip}
20.0 & 3.971 & 3.869 & 3.821 & 3.793 & 3.779  \\
\noalign{\smallskip}

 \multicolumn{6}{c}{Width of electromagnetic transition $1_1^{+} \rightarrow 2_1^+$.}   \\
\noalign{\smallskip}
10.0 & 0.0274 & 0.0280 & 0.0284 & 0.0285 & 0.0286  \\
\noalign{\smallskip}
15.0 & 0.0296 & 0.0290 & 0.0288 & 0.0287 & 0.0286   \\
\noalign{\smallskip}
20.0 & 0.0302 & 0.0294 & 0.0291 & 0.0288 & 0.0287   \\
\noalign{\smallskip}

 \multicolumn{6}{c}{Width of electromagnetic transition $3_1^{+} \rightarrow 2_1^+$.}   \\
\noalign{\smallskip}
10.0 & 0.090 & 0.0929 & 0.0954 & 0.0965 & 0.0966  \\
\noalign{\smallskip}
15.0 & 0.136 & 0.1223 & 0.1140 & 0.1080 & 0.1039   \\
\noalign{\smallskip}
20.0 & 0.156 & 0.1380 & 0.1263 & 0.1183 & 0.1126   \\
\hline\noalign{\smallskip}
\end{tabular*}
\end{flushleft}

\end{multicols}

\begin{table}[htbp]
\caption{The theoretical values of ANCs $A_{th}^{l,J_c}$  and partial decay widths $\Gamma_{th}^{l,J_c}$ of $^8$B states for $^7$Be + p channels obtained for $\hbar \omega$ = 15 MeV and N$^*_{max}$ = 12 input parameters.}\label{table4}
\begin{tabular}{c c c c  c c c c}
\hline\hline\noalign{\smallskip} 
J$^{\pi}$ & E$^{*}_{theor.}$(MeV) & E$_p^{theor.}$(MeV) & E$_p^{exp.}$(MeV) & l($J_c$) & $\Gamma_{th}$($A_{th}$) &$\Gamma_{tot}^{exp}$( $A^{2}_{exp}$)\\
\hline\noalign{\smallskip}
2$_1^+$& 0.0 & -0.151 & -0.137 & 1(1) & 0.463 fm$^{-1/2}$ & 0.452 fm$^{-1}$ \cite{tim-exp}\\

  &  &  &  & 1(2) & 0.623 fm$^{-1/2}$ &  0.711 $\pm$ 0.092 fm$^{-1}$ \cite{tilley-8} \\
  &  &  &  & 3(1) &1.74$\cdot$10$^{-4}$ fm$^{-1/2}$ & \\
  &  &  &  & 3(1) &7.0$\cdot$10$^{-4}$ fm$^{-1/2}$ &  \\
 \hline
1$_1^+$& 1.24 & 1.09 & 0.632 & 1(1) & 15.6 keV & 35.6 $\pm$ 0.6 keV \cite{tilley-8}\\
  &  &  &  & 1(2) &  28.0 keV &  \\
  &  &  &  & 3(2) & 0.327 eV &  \\  
\hline
3$_1^+$& 2.316 & 2.165 & 2.182 & 1(2) & 850 keV & 350 $\pm$ 30 keV \cite{tilley-8}\\
  &  &  &  & 3(1) & 25.0 eV &  \\
  &  &  &  & 3(2) & 173 eV &  \\ 
  
\noalign{\smallskip}\hline\hline
\end{tabular}
\end{table}
  
\begin{multicols}{2}

The asymptotic properties of $^8$B nucleus states were calculated for experimental values of proton channel threshold energies for a variety oscillator parameter $\hbar \omega$ values: 10.0, 12.5, 15.0, 17.5 and 20.0 MeV. The calculated partial decay widths and ANCs for $\hbar \omega$ = 15.0 MeV and N$^*_{max}$ = 12 are presented in Table 4. Known experimental data are included for comparison. 

These results together with electromagnetic transitions widths were used for calculation of $^7$Be(p, $\gamma$)$^8$B reaction cross-section, presented in Figure 1. 

\end{multicols}

\begin{center}
\begin{figure}[htp]
\includegraphics[width=0.9\textwidth]{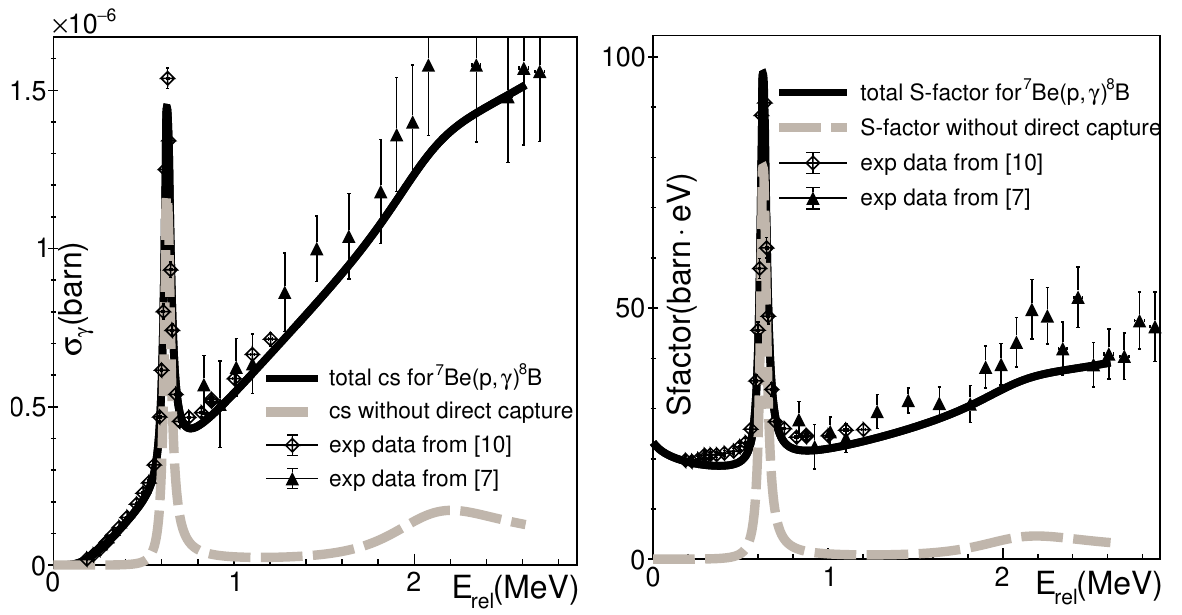}
\caption{The cross-section and S-factor of radiative proton capture by $^7$Be nucleus for $\hbar \omega$ = 15 MeV and N$^*_{max}$ = 12.}\label{fig1}
\end{figure}
\end{center}

\begin{multicols}{2}

As it could be seen from Figure 1 calculations show good agreement with experimental data for both low energies \cite{exp7} and higher energies \cite{exp4}. This good agreement allows us to claim a well-founded derivation of the astrophysical S-factor in energy regions inaccessible to experiment, i. e. energies below 180 keV. 
Figure 2 shows the result of partial analysis of the impact of $^8$B resonances on the values of the cross-section and S-factor in the same energy region.

\end{multicols}

\begin{center}
\begin{figure}[htp]
\includegraphics[width=0.9\textwidth]{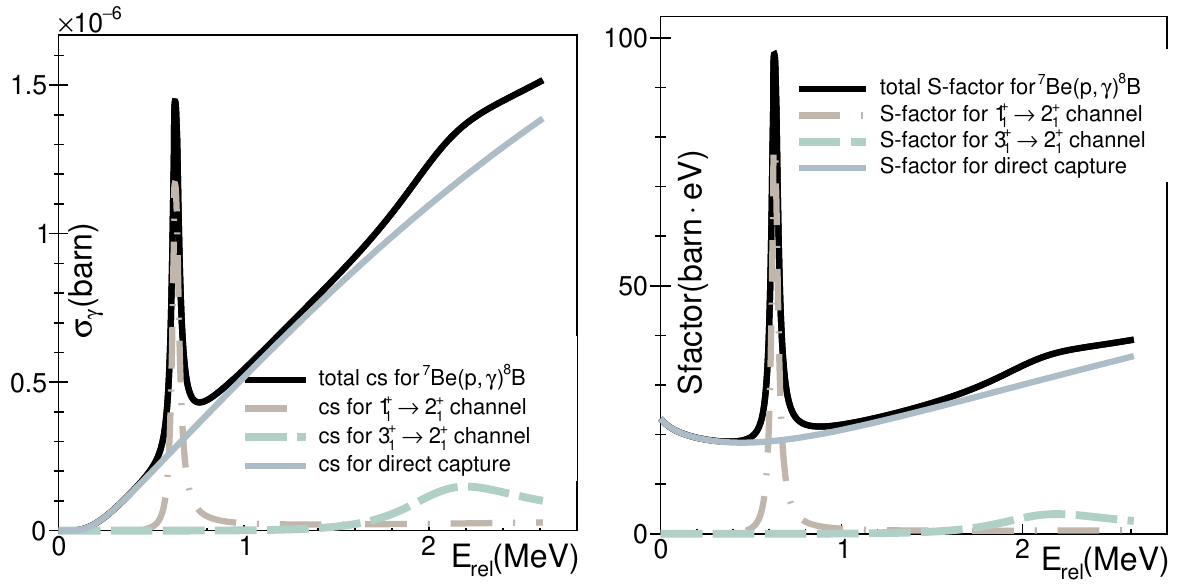}
\caption{The contribution of $^8$B resonances to cross-section and S-factor of radiative proton capture by $^7$Be.}\label{fig1}
\end{figure}
\end{center}

\begin{multicols}{2}
  
As it follows from the analysis of Figure 1, in the energy range up to 2.5 MeV, with the exception of the narrow peak of 1$_1^+$, the cross-section is determined mainly by direct (external) capture to ground 2$_1^+$ state and, consequently, depends, first of all, on ANCs of the decay channels of this state. 

Thus, the total squared value of 2$_1^+$ ANC calculated in this paper is 0.602 fm$^{-1}$. This value is between the value presented in the tables of \cite{tilley-8}, which is 0.711 $\pm$ 0.092 fm$^{-1}$, and the compiled result of 0.452 fm$^{-1}$ contained in \cite{tim-exp}. Anyway, the cross-section obtained using the discussed ANC values turn out to be in good agreement with the measured ones. The results of testing the quality of the performed computations of ANCs are presented below.

It could be seen from Table 4, the total width of the 1$_1^+$ is also in a good agreement with the measured one: 43.6 keV vs 35.6 $\pm$ 0.6 keV. The width of 3$_1^+$ is overestimated compare to known data. However, the experimental decay width of 3$_1^+$ state is obtained mainly by the R-matrix analysis of the cross-section of $^7$Be(p, $\gamma$)$^8$B reaction itself in the energy region around 2 MeV. The measured cross-section in this energy region is characterized by rather large errors \cite{exp4} -- see Figure 1, as well as by the dominating contribution of direct capture amplitude -- see Figure 2. The contribution of 3$_1^+$ resonance to the total cross-section does not exceed 13 \% in its maximum, which is comparable to the measurement accuracy of about 11 \% in this energy area. Thus, an experimental estimate of the 3$_1^+$ resonance width can hardly be called reliable. So that, the values of the width of 3$_1^+$ state obtained in this work can be considered as quite reasonable. A procedure of testing the quality of computations of the decay widths was also carried out (see bellow).

It is important to emphasize that the goal of the study we present in this paper was not only to obtain, on the basis of ab initio calculations, the cross section and the astrophysical S-factor of reaction $^7$Be(p,$\gamma$)$^8$B in the measured and unmeasured energy regions, but also, to no lesser extent, to test the presented method as a whole using an example that has been well studied experimentally with a view to its widespread use.

The most important methodical element of the studies using NCSM, CCOFM, and R-matrix calculations is the analysis of reliability of the results obtained by use of them. Fast convergence of the widths of M1 transitions are demonstrated by Table 3, which allows to get accurate results for these values in NCSM calculations. More difficult question of nuclear size parameters convergence has also been satisfactorily resolved in several ways. In particular, the authors had developed twisted tape extrapolation method - TTE for these purposes \cite{our-radii2}. These methods look the most effective for estimating the probabilities of E2-transitions, but, as shown above, within the framework of the problem under study, any precise calculation of these values is not required to solve the problem under discussion.

The problem of convergence of asymptotic characteristics of closed and open decay channels has not been reliably resolved yet. So, of great interest is the evolution of the ANCs and RPWAs with increasing N$^*_{max}$ in a wide range of $\hbar \omega$ values. In Table 5 the convergence pattern of the results of CCOFM calculations of ANCs and partial widths for basic proton decay channels of $^8$B nucleus is shown.

A visual analysis of the data presented in the table creates a contradictory impression. On the one hand, the columns containing the calculation results on the maximum available basis, characterized by the boundary value of N$^*_{max}$ = 12, in all cases show lines of very high stability for a fairly wide range of $\hbar \omega$ values. On the other hand, the change in the investigated decay widths in the rows of the Table 5 in the range of values of N$^*_{max}$ = 8$\div$12 is not small.

To resolve the discussed contradiction, the following extrapolation procedure was used. First, for each value of the oscillator parameter the set of computed values of $\Gamma^{c_\kappa}_J$ was extrapolated by use of simple two-parameter exponential formula 

\begin{equation}
\Gamma(N^*_{max})=\Gamma(\infty)\cdot (1 - \alpha \cdot e^{-\beta N^*_{max}}),
\label{ext}
\end{equation}
where $\alpha$, $\beta$ are the fitting parameters. In spite of good convergence of ANCs values, the same extrapolation procedure was carried out for them for reliability. The results of the extrapolation are also shown in the table. Taking into account the task of checking the computational accuracy, some of these results contain four significant digits. 

Let us consider the ANC of 2$_1^+$ state decaying to l=1, $J_c$=1 virtual channel $A_{2_{1}^+}^{l=1, J_c=1}$. The set of results of its calculation is a smooth and almost flat horizontal surface above plane $\hbar \omega \times$ N$^*_{max}$ for all studied values of $\hbar \omega$ in the range N$^*_{max}$  = 12$\div $ extrap. The results of the extrapolation procedure confirm with very good accuracy the results obtained for N$^*_{max}$ = 12. By averaging the extrapolated values, one can obtain its mean value and standard deviation equal to 0.464 $\pm$ 0.002. 

Not taking into account the slightest discrepancy in the last line, ANC $A_{2_{1}^+}^{l=1, J_c=2}$ demonstrates completely analogous behaviour. Its mean value and standard deviation are equal to 0.624 $\pm$ 0.003. As a result, the accuracy of calculating the asymptotic normalization coefficients turns out to be close to 0.5\%.

The convergence of the $\Gamma_{1_{1}^+}^{l=1, J_c=1}$ value is slower than that characteristic of ANCs -- a tendency toward its decrease for N$^*_{max} >$12 is clearly visible. Moreover, the values presented in the last row of the table sharply deviate from the smooth systematic pattern of its change. This last circumstance is not difficult to understand, given that with increasing $\hbar \omega$, the size of the region in which the CFF (\ref{eq9}) is correctly described decreases. However, extrapolation in a narrower range of $\hbar \omega$ =10.0$\div$17.5 MeV leads to results that differ by no more than a few units of the fourth significant digit for different $\hbar \omega$. The final mean value and standard deviation are equal to 15.03 $\pm$ 0.03 keV. As a result, using the extrapolation procedure, a uniquely small for nuclear physics calculations standard deviation of 0.2\% was achieved. The average extrapolated value of $\Gamma_{1_{1}^+}^{l=1, J_c=1}$ differs significantly from that obtained in calculations on a limited basis, but in the discussed particular case this does not actually affect the behaviour of the astrophysical S-factor curve due to the dominance of the amplitude of the direct process.

The properties of the data arrays $\Gamma_{1_{1}^+}^{l=1, J_c=2}$ and $\Gamma_{3_{1}^+}^{l=1, J_c=2}$ differ little from those just described. The only difference is that the smooth systematic of the values is violated in the last two rows of the table. Despite this, the accuracy of the results obtained on the narrower space $\hbar \omega \times$ N$^*_{max}$ remains just as uniquely high. Their values are: $\Gamma_{1_{1}^+}^{l=1, J_c=2}$ = 27.11 $\pm 0.09$ keV; $\Gamma_{3_{1}^+}^{l=1, J_c=1}$ = 817.3 $\pm 3.3$ keV. 

Thus, the conducted tests demonstrate that calculations of ANCs and partial widths of single-nucleon decay channels of light nuclei using basis sets accessible to modern computers, supplemented by appropriate extrapolation procedures, provide high numerical accuracy. The origins of discrepancies between these results and reliably measured ones may be conceptual shortcomings of NCSM, CCOFM, and R-matrix theory.

\end{multicols}

\begin{center}
\tabcaption{The results of the CCOFM calculations of ANCs and partial widths for dominating proton capture channels of $^8$B nucleus states.}\label{table5}
\begin{tabular*}{0.985\textwidth}{|c|c c c c c c c c c c c c|}
\hline\hline\noalign{\smallskip} 
 & \multicolumn{4}{c}{$A_{2_{1}^+}^{l=1, J_c=1}$ (fm$^{-1/2}$)} & \multicolumn{4}{|c|}{$A_{2_{1}^+}^{l=1, J_c=2}$ (fm$^{-1/2}$)}  & \multicolumn{4}{c|}{$\Gamma_{1_{1}^+}^{l=1, J_c=1}$ (keV)} \\
\hline\noalign{\smallskip}
$\hbar \omega$ / N$^*_{max}$ & 8 & 10 & 12 & \multicolumn{1}{c|}{extrap.} & 8 & 10 & 12 & \multicolumn{1}{c|}{extrap.} & 8 & 10 & 12 & \multicolumn{1}{c|}{extrap.}  \\
10.0 & 0.489  & 0.476 & 0.468 & \multicolumn{1}{c|}{0.467} & 0.653 & 0.639 & 0.629 & \multicolumn{1}{c|}{0.627} & 16.30  & 15.69  & 15.30  & 15.07 \\
12.5 & 0.482  & 0.471 & 0.464 & \multicolumn{1}{c|}{0.464} & 0.624 & 0.625 & 0.626 & \multicolumn{1}{c|}{0.626} & 16.51  & 15.93  & 15.57  & 15.00 \\
15.0 & 0.481  & 0.471 & 0.463 & \multicolumn{1}{c|}{0.463} & 0.612 & 0.619 & 0.623 & \multicolumn{1}{c|}{0.625} & 16.64  & 16.08  & 15.60  & 15.03 \\
17.5 & 0.483  & 0.473 & 0.463 & \multicolumn{1}{c|}{0.462} & 0.601 & 0.611 & 0.616 & \multicolumn{1}{c|}{0.618} & 16.78  & 16.28  & 15.70  & 15.04 \\
20.0 & 0.476  & 0.476 & 0.467 & \multicolumn{1}{c|}{0.467} & 0.587 & 0.593 & 0.612 & \multicolumn{1}{c|}{0.624} & 16.11  & 16.58  & 16.33  & 15.81 \\
\hline
& \multicolumn{6}{c|}{$\Gamma_{1_{1}^+}^{l=1, J_c=2}$ (keV)} & \multicolumn{6}{c|}{$\Gamma_{3_{1}^+}^{l=1, J_c=2}$ (keV)} \\
\hline\noalign{\smallskip}
$\hbar \omega$ / N$^*_{max}$ & & 8 & 10 & 12 & \multicolumn{2}{c|}{extrap.} & & 8 & 10 & 12 & \multicolumn{2}{c|}{extrap.}  \\
10.0  &  & 32.20  & 30.45  & 29.30 & \multicolumn{2}{c|}{27.08} & & 909.3  & 869.0  & 829.0 & \multicolumn{2}{c|}{815.6} \\
12.5  &  & 29.90  & 29.10  & 28.41 & \multicolumn{2}{c|}{27.21} & & 883.8  & 860.2  & 842.0 & \multicolumn{2}{c|}{815.4} \\
15.0  &  & 29.30  & 28.79  & 28.00 & \multicolumn{2}{c|}{27.04} & & 887.3  & 863.5  & 850.0 & \multicolumn{2}{c|}{821.1} \\
17.5  &  & 28.76  & 28.58  & 27.70 & \multicolumn{2}{c|}{26.17} & & 904.2  & 878.4  & 862.0 & \multicolumn{2}{c|}{833.7}  \\
20.0  &  & 26.90  & 27.97  & 28.00 & \multicolumn{2}{c|}{28.50} & & 899.8  & 914.0  & 870.0 & \multicolumn{2}{c|}{876.7} \\
\hline

\hline\noalign{\smallskip}
\hline\noalign{\smallskip}
\end{tabular*}
\end{center}

\begin{multicols}{2}

Naturally, the most important thing for astrophysical research is to obtain reliable values of the S-factor of $^7$Be(p, $\gamma$)$^8$B process at energies inaccessible to measurements. In this regard, we conducted a test of the accuracy of the calculation results, similar to the one presented above.
The calculated S$_{17}$(0)-factors of the process for N$^*_{max}$=8, 10, 12 and oscillator parameter range $\hbar \omega$ = 10$\div$20 MeV and the extrapolated ones are presented in Table 6. Extrapolation formula (\ref{ext}) was used for these purposes. 

Convergence of the S-factor is not achieved in the N$^*_{max} \leq$12 range. Besides that, the values presented in the last row of the table sharply deviate from the systematic pattern of its change. At the same time, the extrapolated S-factor values lie with high accuracy on a horizontal line within the range of $\hbar \omega = 10\div$17.5 MeV variation. This allows, within the framework of the averaging procedure analogous to that used above, to obtain the value of S$_{17}$(0) and its standard deviation. Finally, the optimal value of the S-factor turned out to be 23.00 $\pm 0.10$ [eV $\cdot$ Barn].

Of interest is also the behaviour of the astrophysical S-factor at proton energies close to zero point. This behaviour is illustrated by Figure 3. As can be seen from this graph, the S-factor decreases rather quickly with the grows of proton energy. For low energies, close to zero point it can be obviously assumed that the standard deviation of S-factor is the same as the standard deviation of S$_{17}$(0).

\begin{table}[H]
\begin{flushleft}
\caption{ The results of the calculations of the astrophysical S-factor S$_{17}$(0) [eV $\cdot$ Barn] of $^7$Be(p, $\gamma$)$^8$B reaction.}
\nopagebreak
\begin{tabular*}{0.45\textwidth}{c c c c c}
\hline\noalign{\smallskip}
$\hbar \omega$ / N$^*_{max}$ & 8 & 10 & 12 & extrap.  \\
\hline\noalign{\smallskip} 
10.0 & 25.68 & 24.49 & 23.66 & 23.16 \\
12.5 & 23.94 & 23.62 & 23.41 & 22.94 \\
15.0 & 23.37 & 23.36 & 23.24 & 22.97 \\
17.5 & 22.94 & 23.09 & 22.97 & 22.94 \\
20.0 & 22.040 & 22.3 & 22.86 & 23.95 \\
\noalign{\smallskip}
 
\hline\noalign{\smallskip}
\end{tabular*}
\end{flushleft}
\end{table}

\begin{center}
\includegraphics[width=1.00\linewidth]{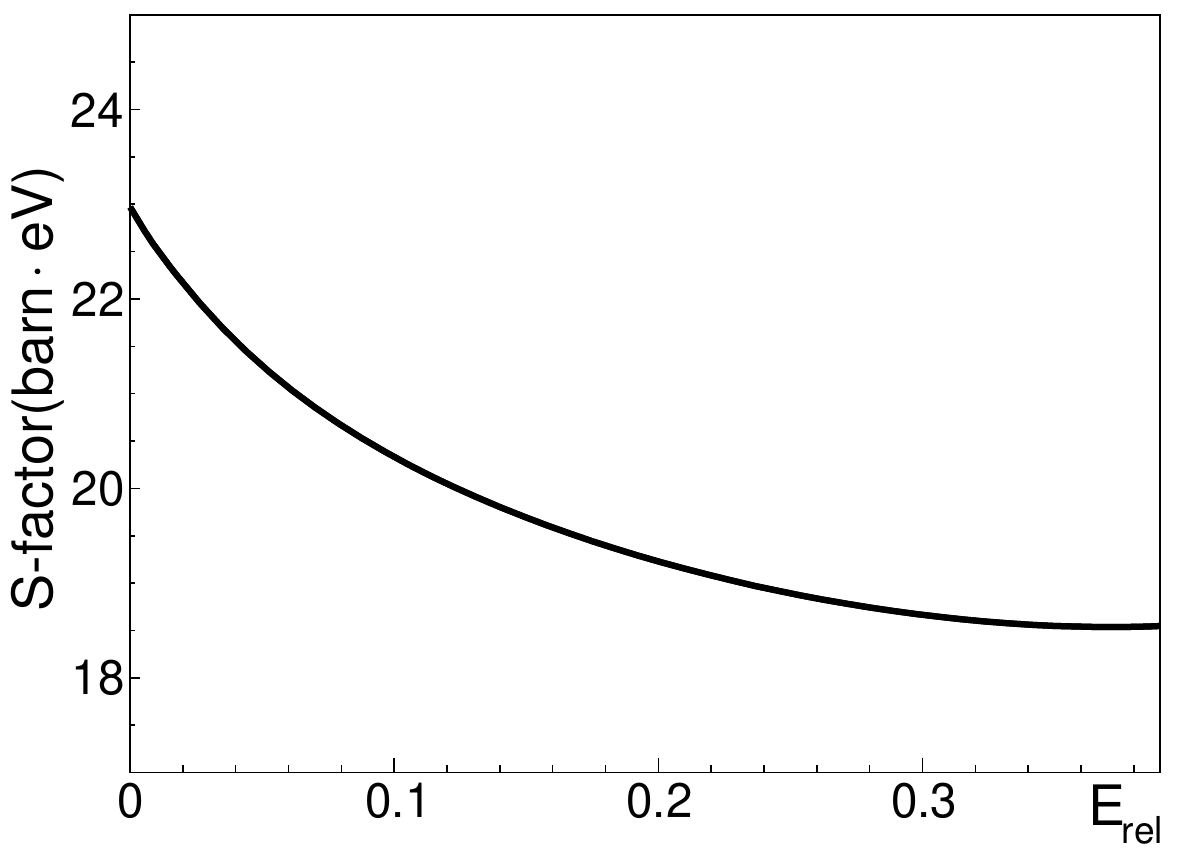}
\figcaption{\label{fig3} Behaviour of S-factor at low energies.}
\end{center}

\section{Conclusions}

In conclusion let us list the basic points of the performed investigation.

\noindent
1. With the use of new approach, based on Cluster Channel Orthogonal Functions Method together with NCSM calculations of the wave functions of $^7$Be and $^8$B states and probabilities of electromagnetic transition between them, ab initio calculations of the astrophysical S-factor of proton radiative capture by $^7$Be nucleus were carried out. 

\noindent
2. The NCSM calculations were carried out with the use of Daejeon16 potential the high efficiency of which has been demonstrated in a large number of studies. The size of the NCSM basis used is almost 10$^9$ SD. Because of the high sensitivity of the results of calculating the asymptotic properties and transition probabilities to the process energy, the measured values of the energy levels and the proton decay threshold were used. The remaining values were calculated ab initio with subsequent use of an extrapolation procedure.

\noindent
3. In this work, for the first time in scientific literature, the question of convergence of NCSM+CCOFM computations of asymptotic properties was raised and resolved. The study of convergence and extrapolation for infinite oscillator basis of S$_{17}(0)$ factor was carried out. Due to that, capabilities of the developed approach turned out to be not limited to the direct calculation of the astrophysical S-factor. The scheme as a whole allows, to perform qualitative analysis of resonance energies, the widths of the electromagnetic transitions and the asymptotic properties of nuclear states in order to identify the amplitudes of various transitions that are significant and, conversely, have little effect on the results of calculations of the astrophysical S-factor at low energies. Moreover, the approach makes it possible to conduct the quantitative analysis of accuracy of the obtained values of electromagnetic transition probabilities, asymptotic normalization coefficients and decay widths of proton channels and astrophysical S-factor itself. 
 
\noindent
4. Almost all theoretically obtained characteristics of the states of the nuclei participating in the reaction under study are in good agreement with known experimental data. The exception are the value of the energy of level 1$_1^+$ and the decay width of state 3$_1^+$. This difference between the calculated and measured energies is within the limits of typical NCSM calculation results. The fact that the value of the width of state 3$_1^+$ is evaluated mainly on the basis of measurements of the cross-section of reaction $^7$Be(p, $\gamma$)$^8$B, which do not have high accuracy and theoretically obtained cross-section of this reaction show good agreement with experimental data for both low energies \cite{exp7} and higher energies \cite{exp4} gives the opportunity to assume that for the width of 3$_1^+$ our results can be considered as more reasonable than the ones tabulated in \cite{tilley-8}. The obtained value of astrophysical S-factor S$_{17}(0)$ is also in rather good agreement with earlier theoretical and experimental works. 
 
\noindent
5. Finally, basing on the obtained results of studying the astrophysical S-factor of $^7$Be(p, $\gamma$)$^8$B reaction and 
taking into account the high versatility and reliability of the Cluster Channels Orthogonal Functions Method in solving problems of computing asymptotic normalization coefficients and decay widths of nuclear states into various channels (including cluster ones), one can with full justification hope that the approach as a whole, developed in this work, will find wide application in studies of problems of nuclear astrophysics.

\vspace{5mm}

\acknowledgments{
The study was supported by Russian Science Foundation grant No 25-22-00651.
}

\end{multicols}

\vspace{10mm}

\begin{multicols}{2}

\end{multicols}

\clearpage
\end{CJK*}

\begin{thebibliography}{90}

\bibitem{pgamma} E. Tel, M. Sahan, I. H. Sarpun, S. G. Okur, EPJ Web of Conferences \textbf{128}, 01005 (2016). 

\bibitem{sk} S. Fukuda et al. (Super-Kamiokande), Phys. Rev. Lett. \textbf{86}, 5651 (2001), arXiv:hep-ex/0103032.

\bibitem{sno} Q. Ahmad et al. (SNO), Phys. Rev. Lett. \textbf{87}, 071301 (2001), arXiv:nucl-ex/0106015.

\bibitem{exp1} Ralph W., Nuclear Physics \textbf{15}, 411-420 (1960).

\bibitem{exp2} P. D. Parker, Phys.Rev. \textbf{150}, 851-856 (1966).

\bibitem{exp3} R. W. Kavanagh, T. A. Tombrello, J. M. Mosher, D. R. Goosman, Bulletin of the American Physical Society Ser.II, \textbf{14}, 1209 (1969).

\bibitem{exp4} F. J. Vaughn, R. A. Chalmers, D. Kohler, L. F. Chase, Phys. Rev. C \textbf{2}, 1657, (1970).

\bibitem{exp5} B. W. Filippone, A. J. Elwyn, C. N. Davids, D. D. Koetke, Phys. Rev. C \textbf{28}, 2222 (1983).

\bibitem{exp6} L.T. Baby et al. (ISOLDE Collaboration),  Phys. Rev. Lett. \textbf{90}, 022501 (2003).

\bibitem{exp7} A. R. Junghans, K. A. Snover, E. C. Mohrmann, E. G. Adelberger, L. Buchmann, Phys. Rev. C \textbf{81}, 012801 (2010).

\bibitem{extrap} R. H. Cyburt, B. Davids, and B. K. Jennings, Phys. Rev. C \textbf{70} 045801 (2004).

\bibitem{breakup} L. Trache, F. Carstoiu, C. A. Gagliardi, A. M. Mukhamedzhanov, R. E. Tribble, Nuclear Physics A \textbf{718}, 493 (2003).

\bibitem{cheft1} A. M. Shirokov, I. J. Shin, Y. Kim, M. Sosonkina, P. Maris, J.
P. Vary, Phys. Lett. B  \textbf{761} 87 (2016).

\bibitem{cheft2} R. Machleidt, D. R. Entem, Phys. Rep. \textbf{503} 1 (2011).

\bibitem{cheft3} D. R. Entem, R. Machleidt, Phys. Rev. C  \textbf{66} 014002 (2002).

\bibitem{jisp1} A. M. Shirokov, J. P. Vary, A. I. Mazur, T. A. Weber, Phys.
Lett. B \textbf{644} 33 (2007).

\bibitem{ncsm1} C. Stump, J. Braun, R. Roth, Phys. Rev. C \textbf{93}, 021301 (2016).

\bibitem{gsm1} G. Papadimitriou, J. Rotureau, N. Michel, M. P. Loszajczak,
B. R. Barrett, Phys. Rev. C \textbf{88}, 044318 (2013).

\bibitem{gfmc1} S. C. Pieper, R. B. Wiringa, Ann. Rev. Nucl. Part. Sci. \textbf{51} 53
(2001).

\bibitem{ccm1} H. Kummela, K. H. Luhrmann, J. Zabolitzky, Phys. Rep. \textbf{36}
1 (1978).

\bibitem{ncsmrgm1} S. Quaglioni, P. Navratil, Phys. Rev. C \textbf{79} 044606 (2009).

\bibitem{ncsmc1} S. Baroni, P. Navratil, S. Quaglioni, Phys. Rev. C \textbf{87} 034326 (2013).

\bibitem{fmd1} T. Neff, H. Feldmeier, Int. J. Mod. Phys. E \textbf{17} 2005 (2008).

\bibitem{gcm1} A. Adahchour, P. Descouvemont, Nucl. Phys. A \textbf{813} 252
(2008).

\bibitem{mcm1} K. Arai, P. Descouvemont, D. Baye, W. Catford, Phys. Rev. C \textbf{68}, 014310 (2003).

\bibitem{amd1} Y. Kanada-Enyo, H. Horiuchi, Prog. Theor. Phys. Suppl. \textbf{142}, 205 (2001).

\bibitem{avrgm1} A.S. Solovyev, S.Y. Igashov, Phys. Rev. C \textbf{96}, 064605 (2017).

\bibitem{cc-gsm1} K. Fossez, N. Michel, M. Ploszajczak, Y. Jaganathen, and R. M. Id Betan, Phys. Rev. C \textbf{91}, 034609 (2015).

\bibitem{hfb1} L. Anh, M. Loc., Phys. Rev. C \textbf{106} 1 014605 (2022).

\bibitem{ccm2} L. Asgari, and H. Sadeghi, Pramana-J. Phys. \textbf{98} 50 (2024). 

\bibitem{ccm3} R. Higa, P. Premarathna, and G. Rupak, arXiv:2010.13003v2.

\bibitem{ncsmc2} P. Navratil, K. Kravvaris, P. Gysbers, C. Hebborn, G. Hupin, and S. Quarglioni, Journal of Physics: Conference Series \textbf{2586} (012062 (2023).

\bibitem{our0} D. M. Rodkin, and Yu. M. Tchuvil'sky, Phys. Rev. C \textbf{103}, 024304 (2021).

\bibitem{our1} D. M. Rodkin, and Yu. M. Tchuvil'sky, Phys. Rev. C \textbf{104}, 044323 (2021).

\bibitem{our2} D. M. Rodkin, and Yu. M. Tchuvil'sky, JETP Letters \textbf{119}, 10, 739 (2024).

\bibitem{our3} D. M. Rodkin, and Yu. M. Tchuvil'sky, Int. J. Mod. Phys. E 33 (11) 2441008 (2024).

\bibitem{our4} D. M. Rodkin, and Yu. M. Tchuvil'sky, Int. J. Mod. Phys. E 33 (11) 2441019 (2024). 

\bibitem{talmi-moshinsky} Yu. F. Smirnov, Nucl. Phys. \textbf{39}, 346 (1962).

\bibitem{nem} O. F. Nemets, V. G. Neudachin, A. E. Rudchik, Yu. F. Smirnov, Yu. M. Tchuvil'sky, {\it Nuclear Clusters in Atomic Nuclei and Multinucleon Transfer Reactions} , (Naukova Dumka, Kiev, 1988).

\bibitem{our5} D. M. Rodkin, Yu. M. Tchuvilsky, Chin. Phys. C \textbf{44},  12410 (2020).

\bibitem{azure} D. Odell, C. Brune, D. Phillips, R. deBoer, S. Paneru, Front. Phys. 10:888476 (2022).

\bibitem{bigstick} Calvin W. Johnson et al., arXiv: 1801.08432 (2018)

\bibitem{A5} I. J. Shin, Y. Kim, P. Maris, J. P. Vary, C. Forssen, J. Rotureau, and N. Michel, J. Phys. G: Nucl. Part. Phys. \textbf{44}, 075103 (2017).

\bibitem{ncsmc-pheno} M. C. Atkinson, K. Kravvaris, S. Quaglioni, P. Navratil, Phys. Lett B \textbf{860}, 139189 (2025).

\bibitem{tilley-8} D. R. Tilley, J. H. Kelley, J. L. Godwin, D. J. Millener,
J. Purcell, C. G. Sheu, and H. R. Weller, Nucl. Phys. A745 \textbf{155} (2004).

\bibitem{tim-exp} N. K. Timofeyuk, Phys. Rev. C 81, 064306 (2010).

\bibitem{our-radii2} D. M. Rodkin, Yu. M. Tchuvilsky, Phys. Rev. C \textbf{106}, 034305 (2022).

\bibitem{cross} P. Maris and J. P. Vary, Int. J. Mod. Phys. E \textbf{22}, 1330016 (2013).
\end{thebibliography}
\end{document}